\documentclass[review,onefignum,onetabnum]{siamart190516}
\usepackage{bm}
\usepackage{graphicx}
\usepackage{epstopdf}
\ifpdf
  \DeclareGraphicsExtensions{.eps,.pdf,.png,.jpg}
\else
  \DeclareGraphicsExtensions{.eps}
\fi

\def\josaa{J. Opt. Soc. Am. A}
\def\ol{Opt. Lett.}
\def\opex{Opt. Expr.}
\def\pra{Phys. Rev. A}
\def\prb{Phys. Rev. B}

\def\prl{Phys. Rev. Lett.}

\headers{Excitation of Bound States in the Continuum}{Lijun Yuan and Ya  Yan Lu} 

\title{Excitation of bound states in the continuum via second harmonic generations\thanks{Submitted to the editors DATE.
\funding{This work was supported by 
the  Technological Research Program of Chongqing
Municipal Education Commission under project no.~KJ1706155, and the Research Grants Council of Hong Kong
Special Administrative Region, China, under project no.~CityU
11304117.}}} 

\author{Lijun Yuan\thanks{College of Mathematics and Statistics,
    Chongqing Technology and Business University, Chongqing, China 
  (\email{ljyuan@ctbu.edu.cn}).}
\and Ya Yan Lu\thanks{Department of Mathematics, City University of
  Hong Kong, Hong Kong
  (\email{mayylu@cityu.edu.hk}).}}

\usepackage{amsopn}

\begin{document}

\nolinenumbers

\maketitle

\begin{abstract}
A bound state in the continuum (BIC) on a periodic structure 
sandwiched between two homogeneous media is a
guided mode with a frequency and a wavenumber such that propagating plane
waves with the same frequency and wavenumber exist in the homogeneous
media. Optical BICs are of significant 
current interest, since they have applications in lasing,
sensing, filtering, switching, and many light emission processes, but
they cannot be excited by incident plane waves when the structure
consists of linear materials. In this paper,  we study the diffraction
of a plane wave by a periodic structure with a second order
nonlinearity, assuming the structure has a BIC and the frequency and wavenumber
of the incident wave are one half of those of the BIC. Based on
a scaling analysis and a perturbation theory, we show that the incident
wave may induce a very strong second harmonic wave dominated by the
BIC, and also a fourth harmonic wave that cannot be ignored. The
perturbation theory reveals that 
the amplitude of the BIC is inversely proportional to a small
parameter depending on the amplitude of the incident wave and the
nonlinear coefficient.  In addition, a system of four nonlinearly
coupled Helmholtz equations (the four-wave model) is proposed to model
the nonlinear process. Numerical solutions of the four-wave model are presented for
a periodic array of circular cylinders and used to validate the
perturbation results. 
\end{abstract}



\section{Introduction}
\label{introduction}

The concept of bound state in the continuum (BIC)  was originally
introduced by 
von Neumann and Wigner for a quantum system \cite{neumann29}. For
classical waves, a BIC is a trapped or guided mode on a structure with
a frequency 
at which radiative waves carrying power to or from infinity also exist.
Mathematically, a BIC corresponds to an eigenvalue in the continuous
spectrum. At the frequency of the BIC,  boundary value problems for
given incident waves from infinity have no uniqueness \cite{bonnet94}.
BICs have been studied for sound waves, linear water waves, and
electromagnetic waves (including light) based on Helmholtz and
Maxwell's equations. They exist on different kind of structures including
 waveguides with local distortions \cite{evans94}, 
waveguides with lateral leaky channels \cite{plotnik11}, and periodic structures
sandwiched between two homogeneous media
\cite{bonnet94,padd00,shipman03,porter05,shipman07,marinica08,hsu13_2,bulgakov14_2,hu15,yuan17_1}. 
Some BICs exist because they have a symmetry mismatch
with radiative waves, and they are often regarded as  symmetry-protected
\cite{bonnet94,evans94,plotnik11,padd00,shipman03,shipman07,hu15}. 
Other BICs seem to decouple with the radiative waves accidentally
\cite{fried85}. 
  On periodic structures sandwiched between two homogeneous media,
  there could be  
 BICs that propagate along the periodic directions
\cite{shipman03,porter05,marinica08,hsu13_2,bulgakov14_2,yuan17_1}.
These BICs are not protected by symmetry in the usual sense, but 
in some cases, they depend strongly on the symmetry 
\cite{yuan17_2,zhen18}. 
Unlike the well-known
guided modes, a propagating BIC on a periodic structure exists above
the lightline, that is, it co-exists with plane waves in the
homogeneous media for 
 the same frequency and a compatible wavevector.  Recently, 
 BICs have attracted much attention in optics due to their interesting
 properties and potential applications \cite{hzsjs16}. 
Since the BICs are decoupled from the radiative waves, they cannot be
excited by incident waves coming from infinity. Bulgakov {\it et
  al.} \cite{bulg15oe} suggested to use the nonlinear optical Kerr
effect \cite{boyd} to excite the BICs. 
 However, with the Kerr effect, the BICs become 
nonlinear \cite{yuan15} and the associated diffraction problem
exhibits complicated optical bistability and symmetry-breaking
phenomena \cite{yuan16,yuan17feb}.

 In this paper, we study the
excitation of BICs based on a second order nonlinear process \cite{boyd}.
The second order nonlinearity gives quadratic nonlinear terms in the
Maxwell's equations, and it is 
responsible for the second harmonic generation (SHG) process, i.e,
waves with frequency $2\omega$ are generated when the incident wave is of 
frequency $\omega$. SHG is the first nonlinear process studied in the field of nonlinear
optics \cite{franken61}. Most existing studies on SHG aim to increase
the percentage of power converted from the incident wave 
to the second harmonic wave \cite{boyd}. Mathematical problems
associated with SHG have  been studied more than two decades ago
\cite{bao94,bao95n,bao97}.  Our objective is different. We aim to generate a wave
field which is dominated by a BIC. 
To excite a BIC on a periodic structure with frequency $\omega_*$, we
illuminate the structure with an incident wave of 
fundamental frequency $\omega = \omega_*/2$. A second harmonic field (of
frequency $2\omega = \omega_*$) is generated due to the second order
nonlinearity. Part of that second harmonic field is 
proportional to the BIC which does not leak out power. Our objective
is to determine the amplitude of that BIC. Using a perturbation
method, we show that the amplitude of the BIC is proportional to
$A/\delta$, where $A$ is the amplitude of the incident wave, $\delta =
\sqrt[3]{A\gamma_1}$ is a dimensionless small parameter,  and 
$\gamma_1$ is the magnitude of a nonlinear coefficient (one element of the second order nonlinear
susceptibility tensor). In addition, we show that the fourth
harmonic wave (with frequency $2\omega_*$) cannot be 
ignored. The amplitude of the BIC can only be corrected determined
when the fourth harmonic is included in a coupled
system. Therefore, the BIC generation process is very 
different from the standard SHG process. To validate the theoretical
results, we present numerical results for a periodic array of
nonlinear circular cylinders.  

The rest of this paper is organized as follows. 
In section~\ref{formulation}, the governing equations, simplified models
and boundary conditions are derived  for 
two-dimensional (2D) structures and the $E$ polarization. 
A perturbation theory including a formula for the coefficient of the
BIC is developed in section~\ref{perturbation}. 
Numerical methods for solving a nonlinear model with 
four frequencies are presented in
section~\ref{NumericalMethods}. Numerical solutions of the nonlinear  
model are compared with the perturbation results in section~\ref{NumericalResults}. 
The paper is concluded with some remarks in section~\ref{Conclusion}.

\section{Problem formulation}
\label{formulation}
We consider a 2D lossless nonlinear non-dispersive periodic structure which is
invariant in the $z$ direction,  periodic in the $y$ direction with
period $L$, and sandwiched between two identical linear homogeneous
media (given in $x< -D$ and $x>D$ respectively for a positive constant
$D$). 
The structure is described by a real dielectric function $\epsilon =
\epsilon(x,y)$ and a real second order nonlinear susceptibility
tensor \cite{boyd}, both of which are periodic in $y$ with period $L$. In
addition, $\epsilon(x,y) = \epsilon_0 > 0$ and the nonlinear 
susceptibility tensor is zero for $|x| > D$. 
For the $E$ polarization,  the electromagnetic fields are independent of $z$
and the only nonzero component of the electric field is its $z$
component, denoted as $E_z$. In that case, the time-domain Maxwell's
equations with a second order nonlinearity can be reduced to the
following scalar equation  \cite{boyd}
\begin{equation}
\label{eq:wave} \frac{\partial^2 E_z}{\partial x^2} + \frac{\partial^2
  E_z}{\partial y^2} - \frac{\epsilon}{c^2} \frac{\partial^2
  E_z}{\partial t^2} - \frac{\gamma}{c^2} \frac{\partial^2 E^2_z}{\partial t^2} = 0,
\end{equation}
where $c$ is the speed of the light in vacuum and
$\gamma=\gamma(x,y)$ is one element of the second order
nonlinear susceptibility tensor. 

For an incident wave  with frequency $\omega$ (fundamental frequency), we can
expand the wave field in harmonic waves as  
 \begin{equation}
\label{eq:harmonic} 
E_z = \mbox{Re} \left( u_1 e^{-i \omega t} + u_2 e^{-2 i \omega t}  + u_3 e^{-3 i \omega t} + u_4 e^{-4 i \omega t}+ \ldots \right),
\end{equation}
where $u_1=u_1(x,y)$ is the complex amplitude of the fundamental
frequency wave,  and
$u_n=u_n(x,y)$ ($n > 
1$) is the complex amplitude of the  $n$-th  order harmonic wave.
Substituting Eq.~\eqref{eq:harmonic} into Eq.~\eqref{eq:wave}, we
obtain the following infinite coupled system of nonlinear equations:
\begin{equation}
\label{eq:general_freq} 
{\cal L}_n u_n = -\frac{n^2 k^2}{2} \gamma \left( \sum_{m=1}^{n}  u_m u_{n-m} +  2 \sum_{m=1}^{\infty} \bar{u}_m u_{m+n}  \right),
\end{equation}
where $u_0 \equiv 0$,  $k = \omega / c$, and
\begin{equation}
  \label{defLn}
  {\cal L}_n  = \frac{\partial^2}{\partial x^2} + \frac{\partial^2}{\partial y^2} 
 + n^2 k^2 \epsilon.
\end{equation}
The incident wave is
assumed to be a plane wave given in the left linear homogeneous
medium:
\begin{equation}
\label{eq:incident}u^{(i)}_1(x,y)= A e^{i (\alpha x +  \beta y)}, \quad x
< -D,
\end{equation}
where $A$ is the amplitude, $(\alpha, \beta)$ is a real wave 
vector, $\alpha > 0 $ and $\alpha^2 +
\beta^2 = k^2 \epsilon_0$. 

In practice, to solve the nonlinear diffraction problem associated
with the above incident wave, one has to truncate the  infinite coupled
system to a finite system.  In standard theories on SHG \cite{bao94,bao95n,bao97}, high
order harmonic waves are weak,  and it is sufficient to keep the
fundamental frequency and
second harmonic waves. Thus, the infinite system is approximated by 
\begin{eqnarray}
\label{eq:TwoWavesModel_1}  
{\cal L}_1  u_1 &= & -k^2 \gamma \bar{u}_1  u_2, \\ 
\label{eq:TwoWavesModel_2} 
{\cal L}_2  u_2 &=&  -2 k^2 \gamma u_1^2.
\end{eqnarray}
We refer the above system, (\ref{eq:TwoWavesModel_1})-(\ref{eq:TwoWavesModel_2}), as the two-wave model. For many cases, the
second harmonic wave is much weaker than the fundamental frequency wave, so  the nonlinear
coupling term in the right hand side of Eq.~\eqref{eq:TwoWavesModel_1}
can be ignored. In that case, $u_1$ satisfies a linear homogeneous
Helmholtz equation and is unaffected by $u_2$, and $u_2$ satisfies
a linear inhomogeneous Helmholtz equation with $u_1^2$
in the right hand side of Eq.~\eqref{eq:TwoWavesModel_2}
as a source term. This is the so-called un-depleted pump
approximation.  

We are interested in the case where the linear periodic structure
(with the same dielectric function $\epsilon$ and a zero nonlinear 
susceptibility tensor) has a BIC $u_*$ with frequency $\omega_*$. 
The BIC satisfies the linear Helmholtz equation 
\begin{equation}
\frac{\partial^2 u_*}{\partial x^2} + \frac{\partial^2 u_*}{\partial 
  y^2} +  k_*^2 \epsilon u_* = 0, 
\end{equation}
where $k_* = \omega_*/c$ and $u_*$ decays to zero exponentially as
$|x| \to  \infty$.  Since the structure is periodic in $y$, the BIC is
a Bloch mode given as 
\begin{equation}
  \label{bicbloch}
  u_*(x,y) = e^{ i \beta_* y} \phi_*(x,y), 
\end{equation}
where $\beta_* \in [-\pi/L, \pi/L]$ is the real Bloch wavenumber, 
 and $\phi_*$
is periodic in $y$ with period $L$. Importantly, the BIC is above the
lightline, that is,  
\begin{equation}
  \label{abovell}
k_* > |\beta_*|/\sqrt{\epsilon_0}.  
\end{equation}
The above condition implies that $\alpha_* = \sqrt{ k_*^2 \epsilon_0 -
  \beta_*^2}$ is real and the plane waves $e^{i ( \beta_* y \pm \alpha_*
  x)}$ can propagate to or from infinity in the surrounding linear media with
dielectric constant $\epsilon_0$.  In order to excite the BIC by the
second order nonlinear effect, we specify the frequency and
wavevector of the incident wave given in Eq.~\eqref{eq:incident} by 
\begin{equation}
  \label{incfre}
\omega =  \frac{\omega_*}{2},  \quad \beta =  \frac{\beta_*}{2}, \quad
\alpha =  \frac{\alpha_*}{2}.
\end{equation}
 
Since the nonlinear coefficient $\gamma$ is very small for
ordinary nonlinear optical materials, and the incident wave is
normally not extremely strong, we are interested in the regime for
which the dimensionless quantity $ A \gamma_1$ is small, where
$\gamma_1 = \max |\gamma(x,y)|$ is magnitude of $\gamma$. In normal
circumstances, 
  this assumption implies that the third and higher harmonics can be
  ignored and the two-wave model is accurate. Our case is different,
  since the second harmonic wave
  contains a BIC which does not lose power to infinity and there is
  no material dissipation. If we
  consider an initial value problem by turning on the incident wave at
  $t=0$, then the amplitude of the BIC in the second harmonic wave is expected to
  increase with time. However, this amplitude can only approach a
  finite limit as $t \to \infty$, since when the second harmonic wave is sufficiently strong, its own SHG
  (generation of fourth harmonic wave from the second harmonic wave) becomes important. In section~\ref{perturbation}, we show
  that it is necessary to keep the fourth harmonic wave in
  the system. This leads 
to the following three-wave model: 
  \begin{eqnarray}
\label{Three1}
{\cal L}_1 u_1 &= & -k^2 \gamma \bar{u}_1u_2, \\ 
\label{Three2} 
{\cal L}_2  u_2 &=&  -2 k^2 \gamma (u_1^2 + 2 \bar{u}_2 u_4), \\
\label{Three3}  
{\cal L}_4  u_4 &=&  -8 k^2 \gamma u_2^2. 
\end{eqnarray}
We can also keep $u_3$ and use the following four-wave model:
  \begin{eqnarray}
\label{eq:FourWavesModel_1}
{\cal L}_1 u_1 &= & -k^2 \gamma \left( \bar{u}_1 u_2 + \bar{u}_2 u_3 + \bar{u}_3 u_4 \right), \\
\label{eq:FourWavesModel_2} 
{\cal L}_2  u_2 &=&  -2 k^2 \gamma \left( u_1^2 + 2 \bar{u}_1 u_3 + 2 \bar{u}_2 u_4 \right), \\
\label{eq:FourWavesModel_3} 
{\cal L}_3  u_3 &=&  -9 k^2 \gamma \left(u_1 u_2 + \bar{u}_1 u_4 \right), \\
\label{eq:FourWavesModel_4} 
{\cal L}_4 u_4 &=&  -8 k^2 \gamma \left( u_2^2 + 2 u_1 u_3 \right). 
\end{eqnarray}

Since the medium for $|x|>D$ is linear, i.e., $\gamma(x,y)=0$ for
$|x|>D$, the equations for $u_n$, both the exact
Eq.~(\ref{eq:general_freq})  and the approximate ones in the three- or four-wave
models, are reduced to linear homogeneous Helmholtz equations for
$|x|>D$. Therefore, we can 
set up transparent boundary conditions at  
$x=\pm D$ following the standard practice \cite{bao95,bao00}. More precisely, for
integer $n \ge 1$, we let 
\begin{equation}
\beta^{(n)}_m = n \beta + \frac{2 m \pi}{L}, \quad \alpha^{(n)}_m =
\left[ n^2 k^2 \epsilon_0  -  (\beta^{(n)}_m )^2 \right]^{1/2}
\end{equation}
for all integers $m$, define  a linear operator $\mathcal{T}_n$ such
that 
\begin{equation}
\mathcal{T}_n e^{i \beta^{(n)}_m y} = i \alpha^{(n)}_m e^{i \beta^{(n)}_m y}, \quad \mbox{for} \quad m = 0, \pm 1, \pm 2, \ldots,
\end{equation}
then the boundary conditions are 
\begin{eqnarray}
\label{BCpD}  
&& \frac{\partial u_n}{\partial x} =  \mathcal{T}_n u_n,
\quad x= D, \quad n \ge 1, \\
\label{BCnD} 
&&  \frac{\partial u_n}{\partial x} =  - \mathcal{T}_n u_n,
  \quad x= -D, \quad n \ge 2.
\end{eqnarray}
Notice that Eq.~\eqref{BCnD} is not valid for $n=1$. 
Since there is an incident
wave of frequency $\omega$ given in the left homogeneous medium, the
boundary condition of $u_1$ at $x=-D$ should be
\begin{equation}
\label{BCu1nD} \frac{\partial u_1}{\partial x} = -
\mathcal{T}_1 u_1 + 2 i A  \alpha e^{i \beta y},  \quad
x = -D. 
\end{equation}

Moreover, since the structure is periodic in $y$, all harmonic waves
are quasi-periodic in $y$, i.e., 
\begin{eqnarray}
\label{quasi1} 
&&  u_n(x,L/2) = e^{i n\beta L} u_n(x,-L/2), \\
\label{quasi2} 
&& \partial_y u_n(x,L/2) = e^{i n \beta L} \partial_y
u_n(x,-L/2). 
\end{eqnarray}
With boundary conditions \eqref{BCpD}-\eqref{quasi2}, 
we can solve the truncated nonlinear systems, such as the 
the four-wave model, in the following rectangular domain 
\begin{equation}
  \label{defOmega}
\Omega = \left\{ (x,y)\ | \  -D<x<   D, \ -L/2 < y < L/2 \right\}.  
\end{equation}

Assuming the nonlinear coefficient $\gamma$ is real, then the two-,
three- and four-wave models are all energy conserving, in the sense that
the power carried by the incident wave equals the total power radiated
out by the fundamental frequency wave and the retained harmonic
waves. For $n\ge 2$,  the power per period radiated out by $u_n$ can
be calculated by integrating the $x$ component of the complex Poynting
vector at $x=\pm D$ with respect to $y$, and up to a constant scaling,
it is 
\begin{equation}
P_n^{\rm out} = \frac{1}{2nk}  \int_{-L/2}^{L/2} \mbox{Im} \left[
  \overline{u}_n \frac{ \partial u_n}{\partial x} \right]_{x=-D}^{x=D} dy.
\end{equation}
The power (per period) carried by the incident wave is 
\begin{equation}
P_1^{\rm inc} = 
\frac{1}{2k}  \int_{-L/2}^{L/2} \mbox{Im} \left[
  \overline{u}^{(i)}_1 \frac{ \partial u_1^{(i)}}{\partial x} \right]_{x=-D} dy =
\frac{ \alpha L |A|^2}{2k}.
\end{equation}
For $u_1$, we have a reflected wave $u_1^{(r)}$ and a transmitted wave
$u_1^{(t)}$ satisfying $u_1 = u_1^{(i)} + u_1^{(r)}$ for $x < -D$ and 
$u_1 = u_1^{(t)}$ for $x> D$. Thus, the power radiated out by the
fundamental frequency wave is 
\begin{equation}
P_1^{\rm out} =  \frac{1}{2k}  \int_{-L/2}^{L/2} \mbox{Im} \left\{
\left[ 
  \overline{u}_1^{(t)} \frac{ \partial u_1^{(t)}} {\partial x}
 \right]_{x=D} 
-   \left[ \overline{u}_1^{(r)} \frac{ \partial u_1^{(r)}}{\partial x}
  \right]_{x=-D}  \right\}  dy.
\end{equation}
The energy conservation property of the four-wave model is
\begin{equation}
\label{energy4}
  P_1^{\rm inc} = \sum_{n=1}^4 P_n^{\rm out}.
\end{equation}
To show this, we realize that 
\[
\sum_{n=1}^4 \frac{1}{n} \int_\Omega \overline{u}_n {\cal L}_n u_n dx dy 
= -   2 k^2 \int_\Omega \gamma \mbox{Re} \left(
\overline{u}_1^2 u_2 + 
3 \overline{u}_1 \overline{u}_2 u_3  + 
4 \overline{u}_1 \overline{u}_3 u_4  + 2 \overline{u}_2^2 u_4  \right) dxdy
\]
is real. Thus, the imaginary part of the left hand side above is zero,
and it can be written as 
\begin{equation}
\sum_{n=1}^4 \frac{1}{n} \int_{-L/2}^{L/2} \mbox{Im} 
\left[ \overline{u}_n \frac{\partial u_n}{\partial x}
\right]_{x=-D}^{x=D} dy = 0.
\end{equation}
After verifying the term for $n=1$ in the left hand side
above is actually $2k(P_1^{\rm out} - P_1^{\rm inc})$,
Eq.~\eqref{energy4} is obtained. 
The ratio $P_n^{\rm out}/P_1^{\rm inc}$ (for $n\ne 1$) is referred to
as the conversion efficiency for the $n$-th harmonic wave.

\section{Perturbation theory}
\label{perturbation}

In this section, we first determine the scaling for the first a few  harmonic
waves from the exact Eq.~\eqref{eq:general_freq}.  For $A$ and
$\gamma_1$ given in section~\ref{formulation}, we let 
$A\gamma_1 = \delta^3$ and assume $\delta$ is small. The cubic for
$\delta$  is introduced for convenience. 
We also
assume that the periodic structure supports a single (i.e., non-degenerate)
BIC with frequency $\omega_*=2\omega$ and Bloch wavenumber
$\beta_*=2\beta$, and there is no BIC with frequency $n \omega$ and Bloch
wavenumber $n \beta$ for  $n\ne 2$. Since the incident wave
has amplitude $A$, $u_1$ should remain at $O(A)$, $u_3$ and $u_4$ should be
smaller, but $u_2$ may contain a large term proportional to the BIC,
because the power converted to the second harmonic wave can accumulate in the
BIC. Note that the equation for $u_2$ is singular, and it does not
have a solution unless the right hand side is orthogonal to the BIC
$u_*$. The right hand side of the exact equation, Eq.~\eqref{eq:general_freq} for $n=2$,
contains the terms $u_1^2$, $\overline{u}_1 u_3$, $\overline{u}_2
u_4$, etc. In general, the orthogonality condition cannot be satisfied by the term
$u_1^2$ alone. Therefore, one other term must have the same order as
$u_1^2$. Since $u_3$ is smaller than $u_1$ (in magnitude), 
$\overline{u}_1 u_3$ cannot balance $u_1^2$. Therefore, 
$\overline{u}_2 u_4$ should have the same order as $u_1^2$, that is 
$\overline{u}_2 u_4 = O(A^2)$. On the other hand, the scaling of $u_4$
can be determined from the dominant term in the right hand side of its own
equation. The right hand side of Eq.~\eqref{eq:general_freq} for $n=4$
contains $u_1 u_3$, $u_2^2$, $\overline{u}_1 u_5$, $\overline{u}_2
u_6$, etc.  For consistency, it is necessary to assume that $u_2^2$ is
the dominant term.  Therefore, we have $u_4 = O(\gamma_1 u_2^2)$.  From this and the condition $\overline{u}_2 u_4 =
O(A^2)$, it is easy to deduce that 
$u_2 = O(A/\delta)$ and $u_4 = O(A \delta)$. To determine the scaling
for $u_3$, we consider the right hand side of its equation, 
conclude that $u_1 u_2$ should be the dominant term, and thus 
$u_3 = O(A \delta^2)$. Similarly, $u_5 = O(A \delta^4)$, $u_6 = O(A
\delta^3)$, etc.
Since $u_4$ is essential to the equation for $u_2$, a minimal
truncated system may include $u_1$, $u_2$ and $u_4$ only. This leads
to the three-wave model \eqref{Three1}-\eqref{Three3}. We prefer to
keep $u_3$ and use the four-wave model \eqref{eq:FourWavesModel_1}-\eqref{eq:FourWavesModel_4}. Some
small terms in the right hand sides of the four-wave model are probably not
important, because they are on the same order as the terms dropped due to
the truncation. For example, the original equation for $u_4$ has a
term $\overline{u}_2 u_6$ which is $O(A^2 \delta^2)$, same as the
retained term $u_1 u_3$. In the equation for $u_3$, the
term $\overline{u}_1 u_4$ is on the same order as the dropped term 
$\overline{u}_2 u_5$. This suggests that the the four-wave model may
be slightly simplified by removing these terms. However, we will not
explore these minor simplifications. 

Next, we use a perturbation method and the four-wave model 
to find the leading coefficient of
the  BIC in the second harmonic wave. 
Writing the nonlinear coefficient as
$\gamma(x,y) = \gamma_1 F(x,y)$ where $F(x,y)=0$ in the linear media
and $\max|F(x,y)|=1$, we expand $u_1$, $u_2$, $u_3$ and $u_3$ as 
\begin{eqnarray}
\label{eq:expansion_u1} u_1 &=& A \left( u_{10} + \delta^2 u_{11} +
                                \dots  \right), \\
\label{eq:expansion_u2} 
u_2 &=& A  \left( \delta^{-1} u_{20}  +
                                \delta u_{21}  +  \delta^3 u_{22}
                                +  \dots  \right), \\
\label{eq:expansion_u3} u_3 &=& A \left( \delta^2 u_{30} + \delta^4
                                u_{31} + \dots\right), \\ 
\label{eq:expansion_u4} u_4 &=& A \left(  \delta u_{40} + \delta^3
                                u_{41} +  \dots \right),
\end{eqnarray}
where  all $u_{nm}$ for
$n \ge 1$ and $m \ge 0$ are dimensionless functions. The leading terms in the
expansions are consistent with the scaling analysis above. 
The functions $u_1$ and $u_3$ contain even powers of $\delta$
only, while $u_2$ and $u_4$ contain odd powers of $\delta$. This is
consistent with the equations. As an example, consider  the term
$\gamma \overline{u}_1 u_2$ in the right hand side of
Eq.~\eqref{eq:FourWavesModel_1} for $u_1$. 
Since $\gamma$ gives rise to $\delta^3$,
$u_1$ consists of even powers of $\delta$ and $u_2$ consists of odd 
powers of $\delta$, the term $\gamma \overline{u}_1 u_2$ can only
contain even powers of $\delta$, and it is consistent with the left
hand side of Eq.~\eqref{eq:FourWavesModel_1}.

For simplicity, we skip the quasi-periodic boundary
conditions for the perturbation terms, since they exactly follow 
Eqs.~\eqref{quasi1} and \eqref{quasi2}.  
Substituting expansions
\eqref{eq:expansion_u1}-\eqref{eq:expansion_u4}  into 
Eqs.~\eqref{eq:FourWavesModel_1}-\eqref{eq:FourWavesModel_4} and 
boundary conditions \eqref{BCpD}-\eqref{BCu1nD}, 
we obtain a sequence of equations for different powers of $\delta$. 
At $O(1/\delta)$, we have 
\begin{eqnarray}
&&  {\cal L}_2 \, u_{20} = 0,    \quad (x,y) \in \Omega, \\
&& \frac{\partial  u_{20}} {\partial x} = \pm \mathcal{T}_2 u_{20} , \quad   x = \pm D. 
\end{eqnarray}
This is the same system satisfied by the BIC $u_*$. Since we assume
the BIC is a single mode, there must be a constant $C_0$, such that 
$u_{20} = C_0 u_*$. 
At $O(1)$, we obtain 
\begin{eqnarray}
\label{u10eq}
&& {\cal L}_1\,  u_{10} = 0,    \quad (x,y) \in \Omega, \\
&&  \frac{ \partial u_{10} }{\partial x} = \mathcal{T}_1 u_{10} , \quad  x = D, \\
\label{u10bcn}
&&  \frac{ \partial u_{10} }{\partial x} = - \mathcal{T}_1 u_{10} + 2 i \alpha e^{i \beta
   y} ,  \quad  x = -D.
\end{eqnarray}
This is simply the system for the linear fundamental frequency
wave. At $O(\delta)$, we 
obtain systems for $u_{21}$ and $u_{40}$. The system for $u_{21}$ is 
\begin{eqnarray}
&& {\cal L}_2\, u_{21} = 0,    \quad (x,y) \in \Omega, \\
&&  \frac{ \partial u_{21} }{\partial x} = \pm \mathcal{T}_2 u_{21} , \quad   x = \pm D. 
\end{eqnarray}
Therefore, $u_{21}$ is also proportional to the BIC $u_*$, that is
$u_{21} = C_1 u_*$ for a constant $C_1$. This allows us to re-write
Eq.~\eqref{eq:expansion_u2} as
\begin{equation}
\label{newexpansionu2}
  u_2 = A  \left[  (C_0/\delta + C_1 \delta) u_* 
      +  \delta^3 u_{22}
                                +  \dots  \right].
\end{equation}
The system for $u_{40}$ is 
\begin{eqnarray}
&& {\cal L}_4\, u_{40} = -8
   k^2 C_0^2 F u^2_*,  \quad (x,y) \in \Omega,  \\  
&&  \frac{ \partial u_{40} }{\partial x} = \pm \mathcal{T}_4 u_{40} ,  \quad  x = \pm D.
\end{eqnarray}
The right hand side is proportional to $C_0^2$. Since $C_0$ is to be
determined, we define a function $\psi$ such that $u_{40} = C_0^2 \psi$,
then $\psi$ satisfies 
\begin{eqnarray}
\label{psieq}
&& {\cal L}_4 \, \psi = -8  k^2 F u^2_*,  \quad (x,y) \in \Omega,   \\ 
\label{psibcD}
&& \frac{\partial \psi}{\partial x} = \pm \mathcal{T}_4 \psi ,   \quad x = \pm D,
\end{eqnarray}
and it can be solved if the BIC $u_*$ is known. 

At $O(\delta^2)$, we obtain a system for $u_{30}$:
\begin{eqnarray}
\label{u30eq}
&& {\cal L}_3 \, u_{30} = -9 
   k^2 C_0 F u_{10} u_*, \quad (x,y) \in \Omega,  \\  
\label{u30bc}
&& \frac{ \partial u_{30}}{\partial x} = \pm \mathcal{T}_3 u_{30} ,  \quad  x = \pm D. 
\end{eqnarray}
It can be solved after $C_0$ is determined first. At $O(\delta^3)$, we
obtain the follow system for $u_{22}$: 
\begin{eqnarray}
\label{eq:u_22}
&& {\cal L}_2 \, u_{22}  = -2
   k^2  F \left( u_{10}^2 + 2 \bar{C}_0 \bar{u}_{*} u_{40} \right),
   \quad (x,y) \in \Omega, \\    
\label{BCu22}
&& \frac{\partial u_{22}}{\partial x} = \pm \mathcal{T}_2 u_{22} ,  \quad  x = \pm D.
\end{eqnarray}
Notice that the differential equations for $u_{nm}$ are in fact valid
on 
\[
\Omega_{\infty} = \left\{ (x,y)\ | \  -\infty < x < +\infty, \ -L/2 < y <   L/2
\right\}.
\]
To find $C_0$, we replace $u_{40}$ in the right hand side of
Eq.~\eqref{eq:u_22} above by $C_0^2
\psi$, multiply that equation 
by $\bar{u}_*$, and integrate on
$\Omega_{\infty}$. This leads to 
\begin{equation}
\label{eq:C2} |C_0|^2 C_0 = - \frac{\int_{\Omega} F u_{10}^2 \bar{u}_*
dxdy} {2 \int_{\Omega}F \bar{u}_*^2 \psi dxdy},
\end{equation}
where the denominator is assumed to be nonzero. 
Since $F(x,y)=0$ for $|x|>D$, the integrals are evaluated on the
rectangular domain $\Omega$. We observe that $C_0$ depends on the BIC
$u_*$, the linear fundamental frequency wave solution $u_{10}$, and a
scaled fourth harmonic wave $\psi$. 
With $C_0$ obtained,  $u_{40} = C_0^2 \psi$ is already known, $u_{30}$
can be solved from Eqs.~\eqref{u30eq}-\eqref{u30bc}, and $u_{22}$ can be
solved from Eqs.~\eqref{eq:u_22}-\eqref{BCu22}. 

The coefficient $C_1$
can be determined from the solvability condition of $u_{23}$. The
governing equation and boundary condition for $u_{23}$ are
\begin{eqnarray}
\label{eq:u23} 
&& 
\mathcal{L}_2 u_{23} = - 4 k^2 F \left( u_{10} u_{11} + \bar{u}_{10}
   u_{30} + \bar{C}_0  \bar{u}_* u_{41} + \bar{C}_1 \bar{u}_* u_{40}\right), \quad  (x,y) \in \Omega ,\\
&& \frac{\partial u_{23}}{\partial x} = \pm \mathcal{T}_2 u_{23} ,  \quad  x = \pm D.
\end{eqnarray}
Therefore, $u_{23}$ is related to $u_*$, $u_{10}$, $u_{40}$, $u_{30}$,
$u_{11}$ and $u_{41}$. It turns out that $u_{11}$ satisfies 
\begin{eqnarray}
&& \mathcal{L}_1 u_{11} = -  k^2 C_0 F \bar{u}_{10} u_* , \quad  (x,y) \in \Omega ,\\
&& \frac{\partial u_{11}}{\partial x} = \pm \mathcal{T}_1 u_{11} ,
   \quad  x = \pm D, 
\end{eqnarray}
and it can be solved without any difficulty. Meanwhile, $u_{41}$ satisfies 
\begin{eqnarray}
&& \mathcal{L}_4 u_{41} = - 16 k^2 C_0 C_1 F  u_*^2, \quad (x,y) \in \Omega ,\\
&& \frac{\partial u_{41}}{\partial x} = \pm \mathcal{T}_4 u_{41} ,
   \quad  x = \pm D,
\end{eqnarray}
therefore, $u_{41} = 2 C_0 C_1 \psi$, and the equation for $u_{23}$ is
reduced to 
\[
\mathcal{L}_2 u_{23} = - 4 k^2 F \left[  u_{10} u_{11} + \bar{u}_{10}
   u_{30} + \left( 2 |C_0|^2 C_1 + C_0^2 \bar{C}_1 \right)  \bar{u}_*
   \psi \right], \quad  (x,y) \in \Omega.
\]
Multiplying the above equation by $\bar{u}_*$ and integrating on
$\Omega_\infty$, we obtain 
\begin{equation}
  2 |C_0|^2 C_1 + C_0^2 \bar{C}_1 
= - \frac{\int_{\Omega} F \bar{u}_* \left( u_{10} u_{11} + \bar{u}_{10} u_{30}
  \right) dxdy }{\int_{\Omega} F \bar{u}_*^2\psi dxdy}. 
\end{equation}
If $C_0 \ne 0$, $C_1$ can solved from the above equation. 

Notice that $C_0$ is chosen so that the system for $u_{22}$ is solvable,
but the solutions are not unique. To have a unique solution for
$u_{22}$, we must impose the solvability condition for $u_{24}$. In general, 
$u_{2m} = C_m u_*   + u_{2m}^{(p)}$, where  $u_{2m}^{(p)}$ is a
particular solution and $C_m$ is a constant. Clearly, $u_{20}^{(p)} =
u_{21}^{(p)} \equiv 0$. For $m \ge 2$, we can choose the particular
solution to be orthogonal to $u_*$, i.e., 
\begin{equation}
\int \overline{u}_* u_{2m}^{(p)} dxdy = 0, 
\end{equation}
where the domain of integration could be $\Omega_\infty$ or some other
domain chosen for its convenience. 
The coefficient $C_m$ can be determined from the solvability condition
of $u_{2,m+2}$. Therefore, we can write $u_2$ as a sum of two
parts. The first part is proportional to the BIC and the 
the second part is orthogonal to the BIC. More precisely, 
we have $u_2 = B u_* +  u_2^{(p)}$, where 
\begin{eqnarray}
B &=& \frac{A}{\delta} \left( C_0 + C_1 \delta^2 + C_2 \delta^4 +
      \dots \right), \\
  u_2^{(p)} &=&  A\delta^3 \left( u_{22}^{(p)} + \delta^2 u_{23}^{(p)} +
\delta^4 u_{24}^{(p)} + \dots \right).
\end{eqnarray}

It should be emphasized that the same equations for $u_{10}$,
$u_{20}$, $u_{21}$, $u_{22}$ and $u_{40}$, and the same formula for
$C_0$ can be obtained using the three-wave model or more accurate models
containing $u_5$ or more harmonic waves. This implies that the
three-wave model can already give the correct leading order terms in
$u_1$, $u_2$ and $u_4$, but the four-wave model is expected to be more
accurate.  

\section{Numerical method}
\label{NumericalMethods}

In this section, we first present an iterative method for solving the 
four-wave model
\eqref{eq:FourWavesModel_1}-\eqref{eq:FourWavesModel_4} for a
general periodic structure, then describe a domain reduction
technique and a pseudospectral method for the special case of a periodic
array of circular cylinders. 

The four-wave model is a coupled nonlinear system that can only be 
solved by an iterative method. Assuming the $j$-th iterations,
$u_n^{(j)}$ for $1 \le n \le 4$,
 are known, we can derive linear 
partial differential equations for the next iterations, $u_n^{(j+1)}$
for $1\le n \le 4$. The nonlinear term $u_l u_m$, for $l$,  $m \in \{1, 2,
3, 4 \}$, can be approximated by 
$ u_l^{(j)}  u_m^{(j)} + u_l^{(j)} (u_m -u_m^{(j)}) 
+  (u_l -u_l^{(j)}) u_m^{(j)}  
= u_l^{(j)} u_m + u_l u_m^{(j)} -u_l^{(j)} u_m^{(j)}$, 
and it gives rise to 
$u_l^{(j)} u_m^{(j+1)} + u_l^{(j+1)} u_m^{(j)} -u_l^{(j)} u_m^{(j)}$ 
in the equations for the $(j+1)$-th iterations. The nonlinear term 
$u_l \overline{u}_m$
 can be similarly approximated, but the appearance 
of $\overline{u}_m^{(j+1)}$ is inconvenient, since it means that the 
real and imaginary parts of $u_m^{(j+1)}$ must be separated and solved 
in a larger coupled system. We prefer to avoid the complex conjugate 
of $u_m^{(j+1)}$ and simply approximate 
$u_l \overline{u}_m$ by $u_l^{(j+1)} \overline{u}_m^{(j)}$. Therefore,
our iterative method for the four-wave model is 
\begin{eqnarray}
\label{iterative1}
&& {\cal L}_1 u_1^{(j+1)}  + k^2 \gamma \left[ \bar{u}^{(j)}_1 u^{(j+1)}_2 +  
\bar{u}^{(j)}_2  u^{(j+1)}_3 +  \bar{u}^{(j)}_3 u^{(j+1)}_4  \right] = 0, \\
&& {\cal L}_2 u^{(j+1)}_2 + 4 k^2 \gamma \left[   u^{(j)}_1 
   u^{(j+1)}_1 +   \bar{u}^{(j)}_1 u^{(j+1)}_3 + 
   \bar{u}^{(j)}_2  u^{(j+1)}_4  \right] = 2 k^2 
   \gamma [ u^{(j)}_1 ]^2, \\ 
&&  {\cal L}_3 u^{(j+1)}_3 + 9 k^2 \gamma \left[
   u^{(j)}_2 u^{(j+1)}_1  + u^{(j)}_1 
   u^{(j+1)}_2 +  \bar{u}^{(j)}_1 u^{(j+1)}_4  \right] =  9 k^2 \gamma 
   u^{(j)}_1  u^{(j)}_2, \\
\label{iterative4}
&&  {\cal L}_4 u^{(j+1)}_4 + 16 k^2 \gamma \left[ u^{(j)}_3 
   u^{(j+1)}_1 + 
   u^{(j)}_2 u^{(j+1)}_2 +   u^{(j)}_1  u^{(j+1)}_3  \right]  \\
&& \nonumber 
\hspace{5cm} = 8 k^2 \gamma \left\{ [u_2^{(j)}]^2 + 2 
   u_1^{(j)} u_3^{j} \right\}.                                    
\end{eqnarray}  
The above equations are valid on domain $\Omega$ given in 
\eqref{defOmega}. Since the boundary conditions 
\eqref{BCpD}-\eqref{quasi2} are linear, they are directly applicable 
to $u_n^{(j+1)}$. 

Next, we consider a periodic array of circular cylinders with
radius $a$ and dielectric constant $\epsilon_1$, surrounded by a
linear homogeneous medium with  dielectric constant $\epsilon_0$.
The cylinders are infinitely long, identical and parallel to the $z$
axis. Their centers are located on the $y$ axis at $y=mL$ for all  
integers $m$, where $L$ is the period. 
If the cylinders are linear and lossless, the array may support a number
of BICs, including anti-symmetric standing waves
and propagating BICs with a nonzero Bloch wavenumber
\cite{bulgakov14_2,yuan17_1}.  If the cylinders are nonlinear with a
nonzero coefficient  $\gamma_1$ related to the second order nonlinear
susceptibility tensor, the BICs are not solutions to the
nonlinear systems described in section~\ref{formulation}, but an incident wave
can produce a field which is dominated by a BIC.  

For such a periodic array, we can let $D=L/2$, then $\Omega$ is a
square of side length $L$. However, since the medium
outside the cylinders is linear, it is possible to avoid computing different
iterations of the solutions outside the cylinders by finding
boundary conditions for $u_n$ on the boundary of
\begin{equation}
  \label{defdisk}
\Omega_c = \left\{  (x,y) \ | \   x^2 + y^2 < a^2 \right\},
\end{equation}
where $\Omega_c$ corresponds to the cross section of the cylinder
centered at the origin. 
The boundary conditions are given as
\begin{eqnarray}
  \label{bcdisk1} 
  \frac{\partial u_1}{\partial r} &=& {\cal B}_1 u_1 + A f, \quad \mbox{on}
  \quad \partial \Omega_c, \\
\label{bcdiskn}
  \frac{\partial u_n}{\partial r} &=& {\cal B}_n u_n,  \quad \mbox{on}
  \quad \partial \Omega_c \quad \mbox{for} \quad n\ge 2, 
\end{eqnarray}
where $r := \sqrt{x^2 + y^2}$, ${\cal B}_n$ ($n \ge 1$) are operators
acting on functions defined on the circle $r=a$, $A$ is the amplitude
of the incident wave, and $f$ is a function defined on the circle. A
numerical scheme for computing 
${\cal B}_n$ and $f$ is presented in \cite{yuan15}. 
Since there is only an incident wave of frequency $\omega$,  the boundary
conditions for $u_1$ is inhomogeneous, and the boundary conditions for all
$u_n$ ($n\ge 2$) are homogeneous. 
With conditions \eqref{bcdisk1}-\eqref{bcdiskn}, we can solve the
four-wave model in $\Omega_c$ using the iterative 
scheme \eqref{iterative1}-\eqref{iterative4}. Since these boundary
conditions are linear, they are
directly applicable to $u_n^{(j+1)}$ for $1\le n \le 4$. 

Equations \eqref{iterative1}-\eqref{iterative4} can be discretized by
a mixed Chebyshev-Fourier pseudospectral method
\cite{tref,yuan15}. The radial 
variable $r$ is discretized by $M$ points based an extension of $r$ to $[-a,
a]$ and using scaled Chebyshev points for interval $[-a, a]$. The angular
variable $\theta$ (of the polar coordinate system) is uniformly
sampled by $N$ points. 
The differential equations are enforced at $N(M-1)$ discretization
points inside $\Omega_c$. The boundary conditions \eqref{bcdisk1} and
\eqref{bcdiskn} are discretized at $N$ points on $\partial
\Omega_c$. This leads to the following linear system
\begin{equation}
  \label{finalF}
{\bm A}^{(j)}  {\bm u}^{(j+1)} = {\bm b}^{(j)}, 
\end{equation}
where ${\bm u}^{(j+1)}$ is a column vector with four blocks ${\bm
  u}_n^{(j+1)}$ for $1\le n \le 4$, ${\bm u}_n^{(j+1)}$ is a column
vector of length $MN$ representing
$u_n^{(j+1)}$ at the $NM$ discretization points,
${\bm A}^{(j)}$ is a $(4MN)\times (4NM)$ square matrix related to 
${\bm u}^{(j)}$ and its complex conjugate, ${\bm b}^{(j)}$ is a column
vector with four blocks ${\bm b}_n^{(j)}$ for $1\le n \le 4$ and it is 
related to the right hand sides of
\eqref{iterative1}-\eqref{iterative4} and the incident 
wave through the inhomogeneous term in boundary condition
\eqref{bcdisk1}.  Let ${\bm r}^{(j+1)} = {\bm A}^{(j+1)} {\bm
  u}^{(j+1)} - {\bm b}^{(j+1)}$ be the residual vector, and write it
in four blocks ${\bm r}^{(j+1)}_n$ for $1\le n \le 4$, 
we stop the iteration when 
$|| {\bm r}_n^{(j+1)}||/||{\bm b}_n^{(j+1)}||$ and 
$|| {\bm u}_n^{(j+1)} - {\bm u}_n^{(j)} ||/ || {\bm
  u}_n^{(j+1)}||$ for $1\le n \le 4$ are all smaller than an error
tolerance $\tau_{\rm err}$, where 
$||\cdot ||$ is the standard Euclidean norm. 

The perturbation theory of section~\ref{perturbation} gives rise to a
sequence of linear Helmholtz equations. To compute the leading order
terms, we need to calculate the BIC $u_*$, solve $u_{10}$ from 
Eq.~\eqref{u10eq},  solve $\psi$
from Eq.~\eqref{psieq}, evaluate $C_0$ from Eq.~\eqref{eq:C2}, and solve $u_{30}$ from
Eq.~\eqref{u30eq}. Notice that $u_{10}$ is the solution 
of the linear diffraction problem for an incident wave with a unit
amplitude. For a periodic array of circular cylinders, it can be easily solved by a
semi-analytic method based on cylindrical wave expansions in $\Omega$
\cite{yuexia06,martin}. The BIC $u_*$ can also be calculated using the same 
cylindrical wave expansions \cite{hu15,yuan17_1}. 
Since $F(x,y)=0$ for $(x,y)$ outside $\Omega_c$, Eq.~\eqref{psieq} 
is only inhomogeneous in $\Omega_c$, and $\psi$ satisfies the same boundary 
condition as $u_4$ on $\partial \Omega_c$, i.e., 
\begin{equation}
  \label{psibcr}
\frac{\partial \psi}{\partial r} = {\cal B}_4 \psi, \quad 
\mbox{on}\quad \partial \Omega_c. 
\end{equation}
Therefore, Eq.~\eqref{psieq} can be solved in $\Omega_c$ with the above 
boundary condition by the  mixed Chebyshev-Fourier pseudospectral 
method described above. Similarly, $u_{30}$ satisfies
Eq.~\eqref{u30eq} in $\Omega_c$ and the boundary condition  
\begin{equation}
  \label{u30bcr}
\frac{ \partial u_{30} }{\partial r} = {\cal B}_3 u_{30} \quad 
\mbox{on} \quad \partial \Omega_c, 
\end{equation}
and it can be solved by the same pseudospectral method.

\section{Numerical and perturbation results}
\label{NumericalResults}

In this section, we present numerical results for the four-wave model,
validate the perturbation theory  of section~\ref{perturbation}, and
discuss some aspects of the nonlinear process. 


We consider a periodic array of circular cylinders with radius $a =
0.3L$, dielectric constant $\epsilon_1 = 10$, and nonlinear coefficient
$\gamma_1 = 10^{-12}$\,m/V, and assume the surrounding medium
is vacuum with  $\epsilon_0 = 1$. The array has a propagating BIC with angular frequency $\omega_* =
 0.6173\,(2\pi c/L)$ and Bloch
wavenumber $\beta_* = 0.2206\, (2 \pi/L)$. 
We normalize the electric field $u_*$ of the BIC such that  
\begin{equation}
\label{eq:normalization} \frac{1}{L^2} \int_{\Omega_c} |u_*|^2 dxdy = 1.
\end{equation}
Since the periodic array is symmetric in $y$, the BIC can be scaled by a
complex number of unit magnitude such
that $\bar{u}_*(x,y) = u_*(x,-y)$ \cite{yuan17_2}. In addition, we
assume that the phase of $u_*$ with the largest absolute
value in the top half cylinder is non-negative. 
The imaginary and real
parts of the electric field of this BIC are shown in
\begin{figure}[htp]
\centering 
\includegraphics[scale=0.55]{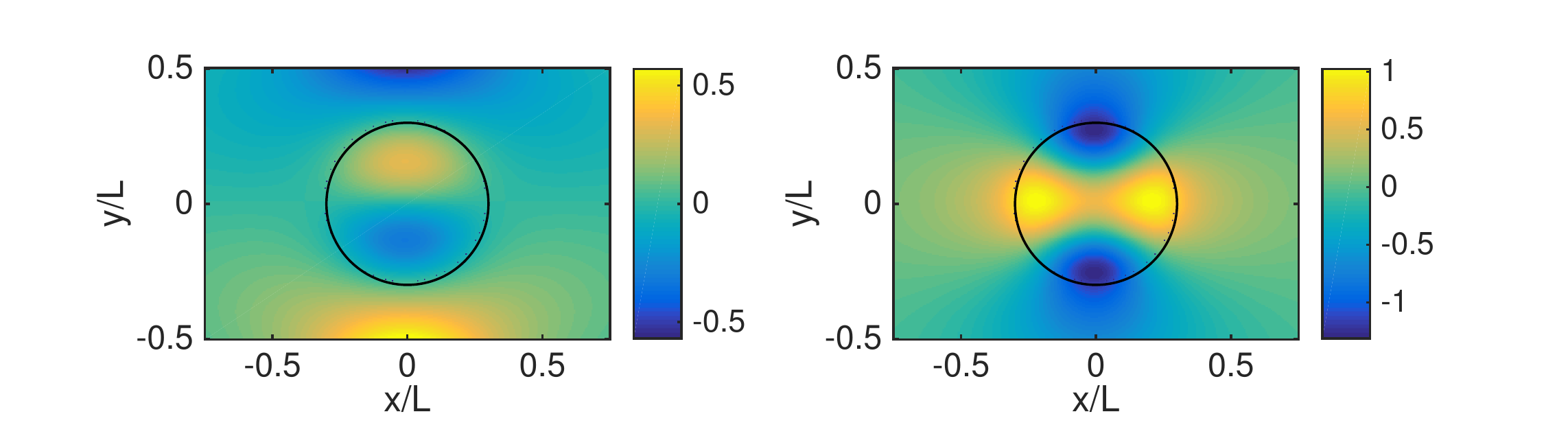}
\caption{The imaginary (left) and real (right) parts of a normalized
  propagating BIC with $\omega_* = 0.6173 (2\pi c/L) $ and $\beta_* 
  = 0.2206 (2\pi/L)$.} 
\label{fig:Example1_BIC}
\end{figure}
Fig.~\ref{fig:Example1_BIC}. 
Notice that $u_*$ is even in $x$. The maximum of $|u_*|$ is about
$1.3165$.   

\begin{figure}[htp]
\centering 
\includegraphics[scale=0.6]{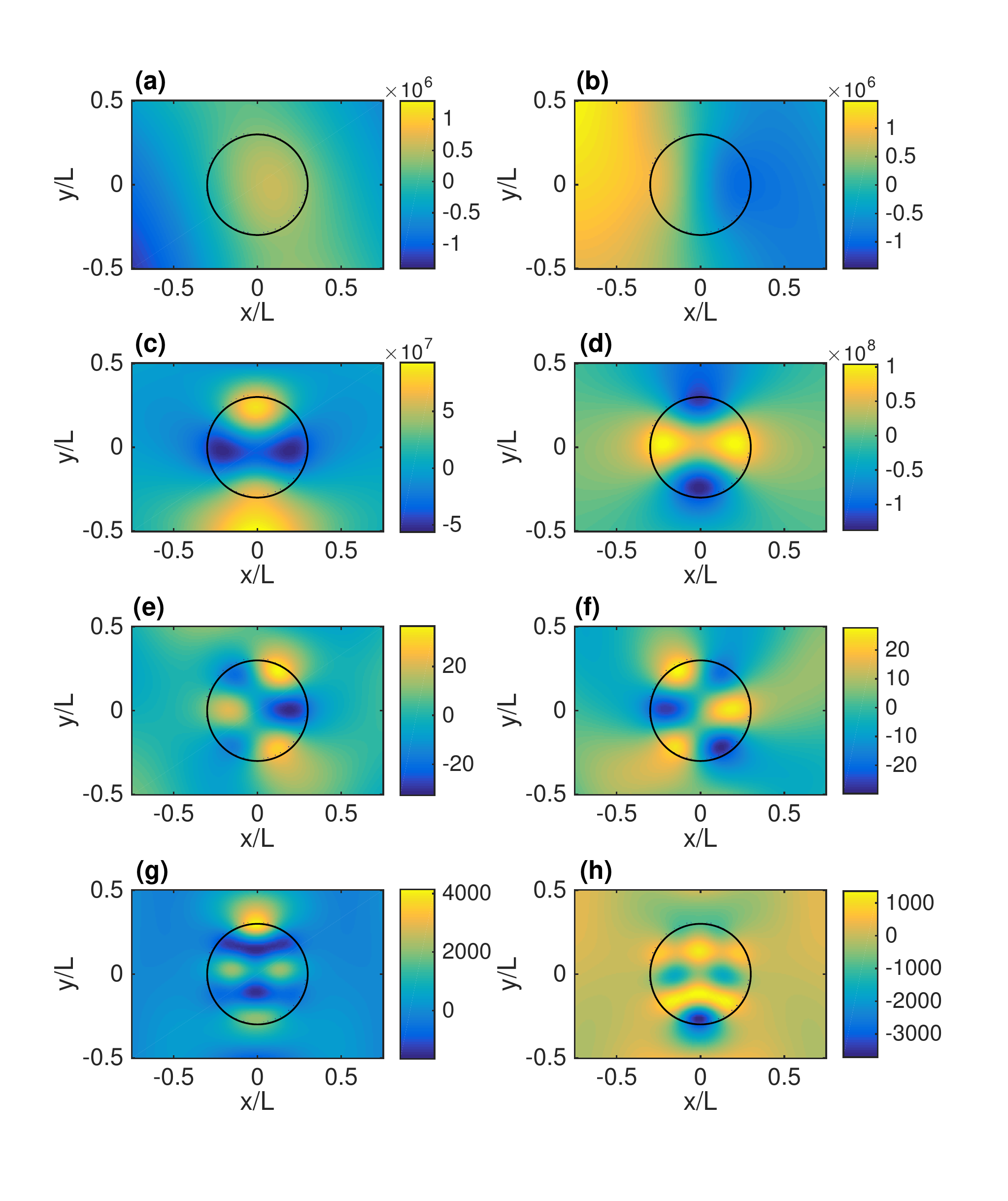}
\caption{Numerical solution of the four-wave model. 
The 
  imaginary and real parts of the 
  fundamental frequency [(a) and (b)], second harmonic [(c) and 
  (d)], third harmonic [(e) and (f)], and fourth harmonic [(g) and 
  (h)] waves are shown in the left and right columns, respectively.}
\label{fig:Example1_Iterative}
\end{figure}

For the nonlinear problem, we specify a plane incident plane wave 
with $\omega  = \omega_*/2 = 0.30865\,(2 \pi c/L) $ and $\beta  =
\beta_*/2 = 0.1103\,(2 \pi/L)$, and assume the  
amplitude of the incident wave  is $A = 10^6$\,V/m. 
We solve the four-wave model by the iterative
scheme \eqref{iterative1}-\eqref{iterative4},  with zero initial
guesses and an error tolerance $\tau_{\rm err} = 10^{-10}$. The results
are shown in Fig.~\ref{fig:Example1_Iterative}.
It can be seen that the second harmonic wave is indeed much stronger
than the incident wave. 
The maximum of  $|u_2|$ is about  $1.4889 \times 10^8$\,V/m.  
This is very different from existing studies on SHG, where
the second harmonic wave is usually very weak when $|A \gamma_1| \ll 1$. 
It can also be seen that the fourth harmonic wave is stronger than the third
harmonic wave, and both of them are weaker than with the incident
wave.  These numerical results are obtained with $N=52$ and $M=26$. 

\begin{figure}[htp]
\centering 
\includegraphics[scale=0.6]{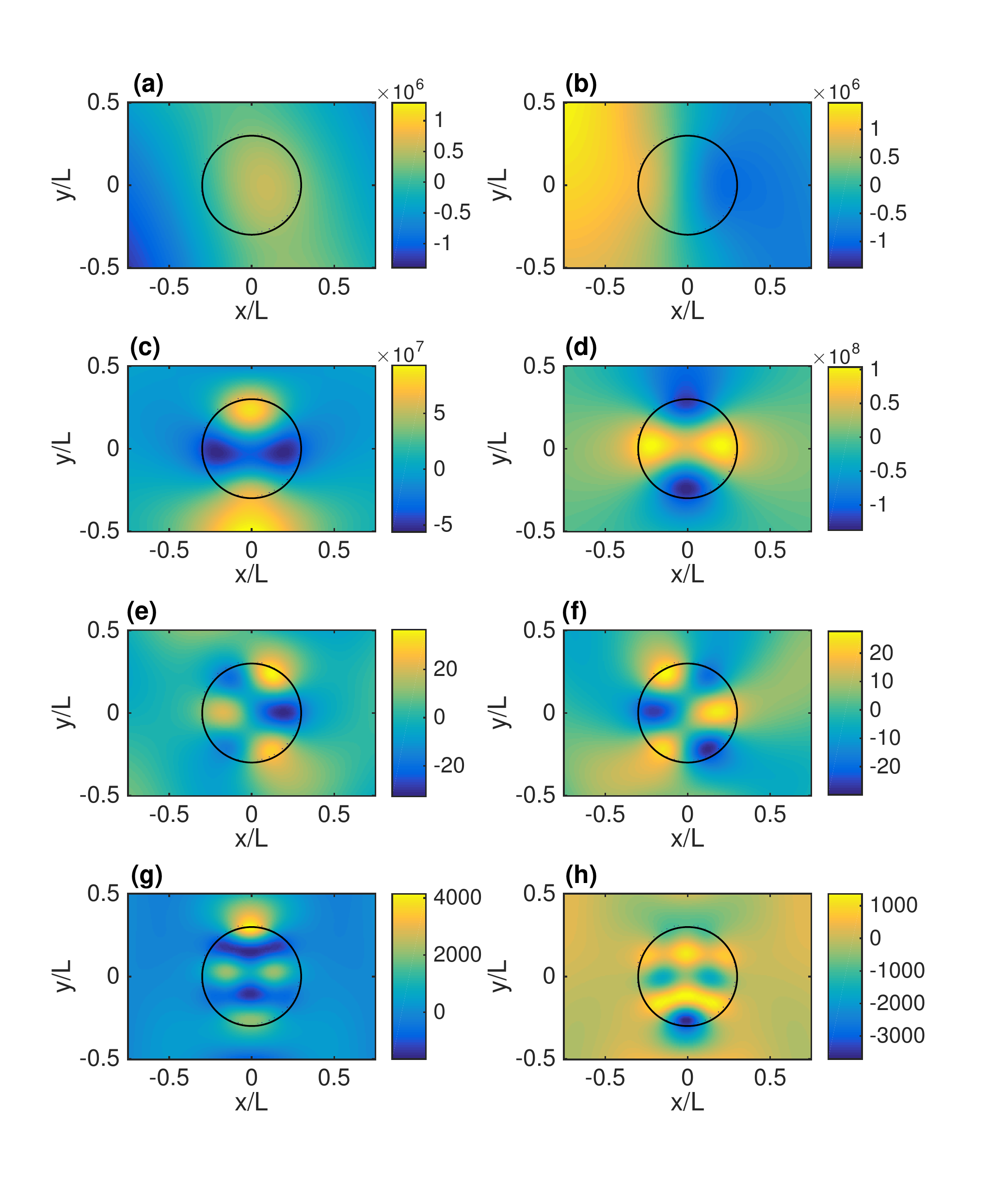}
\caption{Leading terms in the perturbation expansions
  \eqref{eq:expansion_u1}-\eqref{eq:expansion_u4}. 
The   imaginary and real parts of the 
  fundamental frequency [(a) and (b)], second harmonic [(c) and 
  (d)], third harmonic [(e) and (f)], and fourth harmonic [(g) and 
  (h)] waves are shown in the left and right columns, respectively.}
\label{fig:Example1_Perturbation}
\end{figure}

Since $\delta = \sqrt[3]{A \gamma_1} = 0.01$, the numerical
results are consistent with the scaling suggested in
section~\ref{perturbation}, i.e., $u_1 = O(A)$, $u_2 = 
O(A/\delta)$, $u_3 = O( A \delta^2)$ and $u_4 = O(A \delta)$. To check
the validity of the 
perturbation theory, we calculate the leading terms in 
expansions \eqref{eq:expansion_u1}-\eqref{eq:expansion_u4}. 
We solve $u_{10}$ based on 
cylindrical wave expansions in $\Omega$ \cite{yuexia06}, and solve
$\psi$ and $u_{30}$ in $\Omega_c$ by the mixed Chebyshev-Fourier
pseudospectral method. The coefficient $C_0$ given in Eq.~\eqref{eq:C2}
is evaluated, and we get $C_0 = 1.0043 - 0.5201 i$. 
In Fig.~\ref{fig:Example1_Perturbation},
we show the leading terms $A u_{10}$, $(AC_0/\delta) u_*$, $A \delta^2
u_{30}$ and $A \delta u_{40}$ for $u_1$, $u_2$, $u_3$ and $u_4$,
respectively. 
It is clear that they agree very well with the full numerical solutions 
shown in Fig.~\ref{fig:Example1_Iterative}. 
In particular, using the values of $C_0$ and $\max |u_*|$, we get $|AC_0/\delta| \max |u_*| = 1.4889  \times 10^8$\,V/m, and
it agrees perfectly with value of $\max |u_2|$ found earlier. 

\begin{figure}[htp]
\centering 
\includegraphics[scale=0.5]{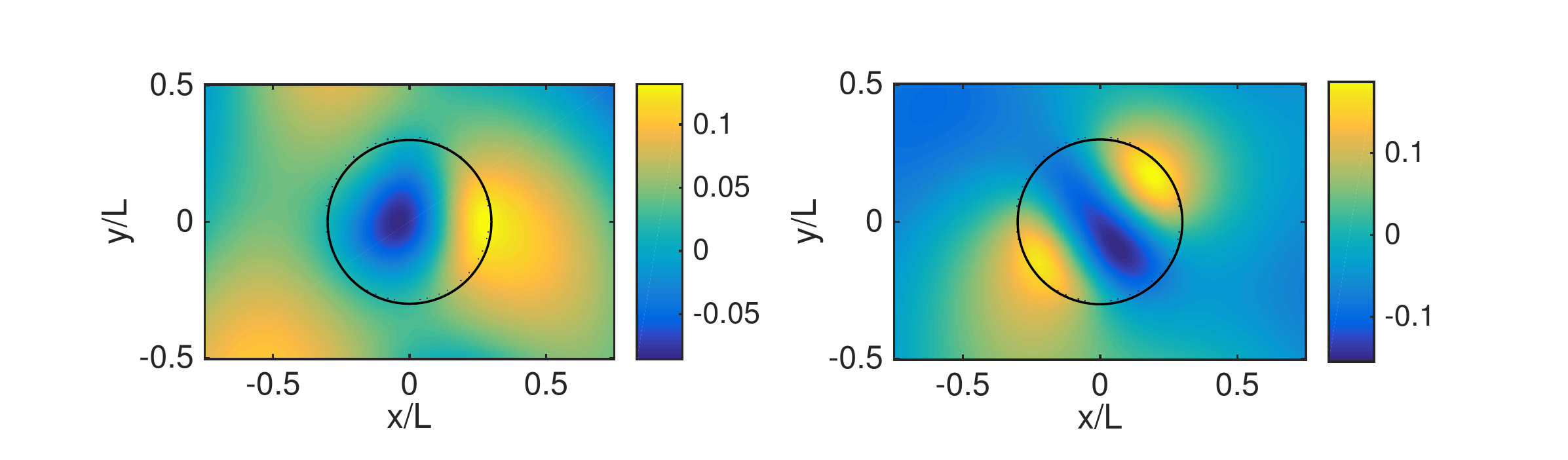}
\caption{The imaginary (left) and real (right) parts of 
$A \delta^3 u^{(p)}_{22}$, where $u^{(p)}_{22}$ is the part of
$u_{22}$ orthogonal to the BIC, $u_{22}$ is the first non-BIC
  term in the perturbation expansion \eqref{eq:expansion_u2} of  the 
  second harmonic wave.}
\label{fig:Example1_2ndTerm_SH}
\end{figure}

For this example, we also calculate the power radiated out by waves 
of different frequencies based on the numerical solutions of the 
four-wave model. 
The energy conservation property $\sum_{n=1}^4  P_n^{\rm
    out} / P_1^{\rm inc}  = 1$ is satisfied with a precision of 
    $10^{-8}$. Although the second harmonic wave $u_2$ is
    strong, very little power is converted to it, since 
the dominate part of $u_2$ is the BIC which cannot radiate out power. 
Among the three harmonic waves, the fourth
harmonic wave has the largest conversion efficiency, followed by the
third harmonic wave and then the second harmonic wave.   
In the perturbation expansion \eqref{eq:expansion_u2} of $u_2$, the first two
terms $u_{20}$ and $u_{21}$ are proportional to the BIC $u_*$, and
only the third term $u_{22}$ can radiate out power. With $C_0$
determined such that the system \eqref{eq:u_22}-\eqref{BCu22} has
solutions, we can calculate $u_{22}$ using the same mixed
Chebyshev-Fourier pseudospectral method. As we mentioned in
section~\ref{perturbation}, 
a unique solution for $u_{22}$ can be determined when 
the solvability 
condition for $u_{24}$ is imposed, and it can be written as 
$u_{22} = C_2 u_* + u_{22}^{(p)}$ with $u_{22}^{(p)}$ orthogonal to
$u_*$ on a chosen domain. 
In Fig.~\ref{fig:Example1_2ndTerm_SH}, 
we show $A \delta^3 u_{22}^{(p)}$ with the orthogonality defined on
$\Omega_c$. The conversion efficiency for $u_2$ is
small, since the magnitude of $A \delta^3 u_{22}^{(p)}$ is much smaller than
those of $u_4$ and $u_3$.

\begin{figure}[htp]
\centering 
\includegraphics[scale=0.4]{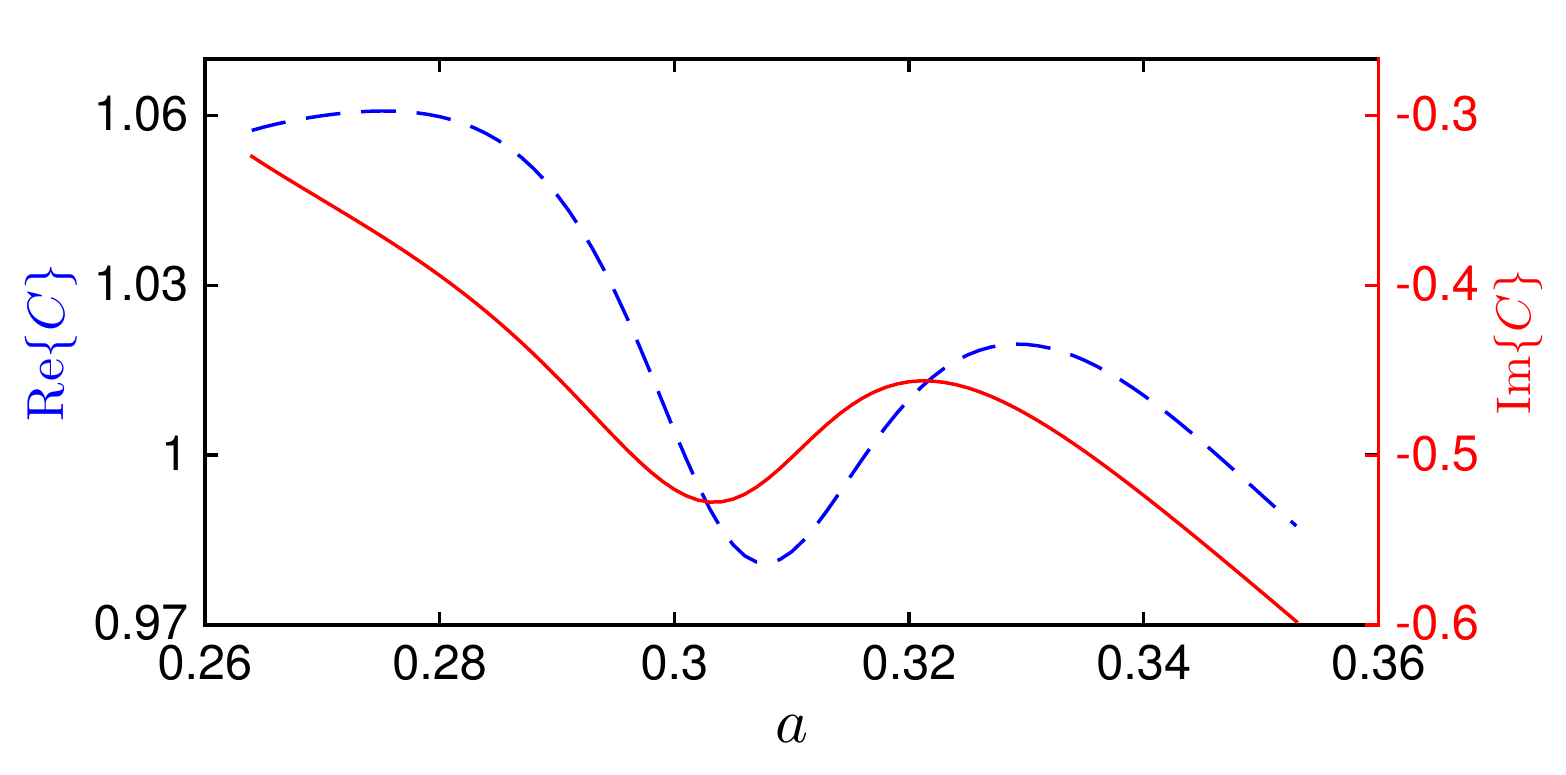}
\includegraphics[scale=0.4]{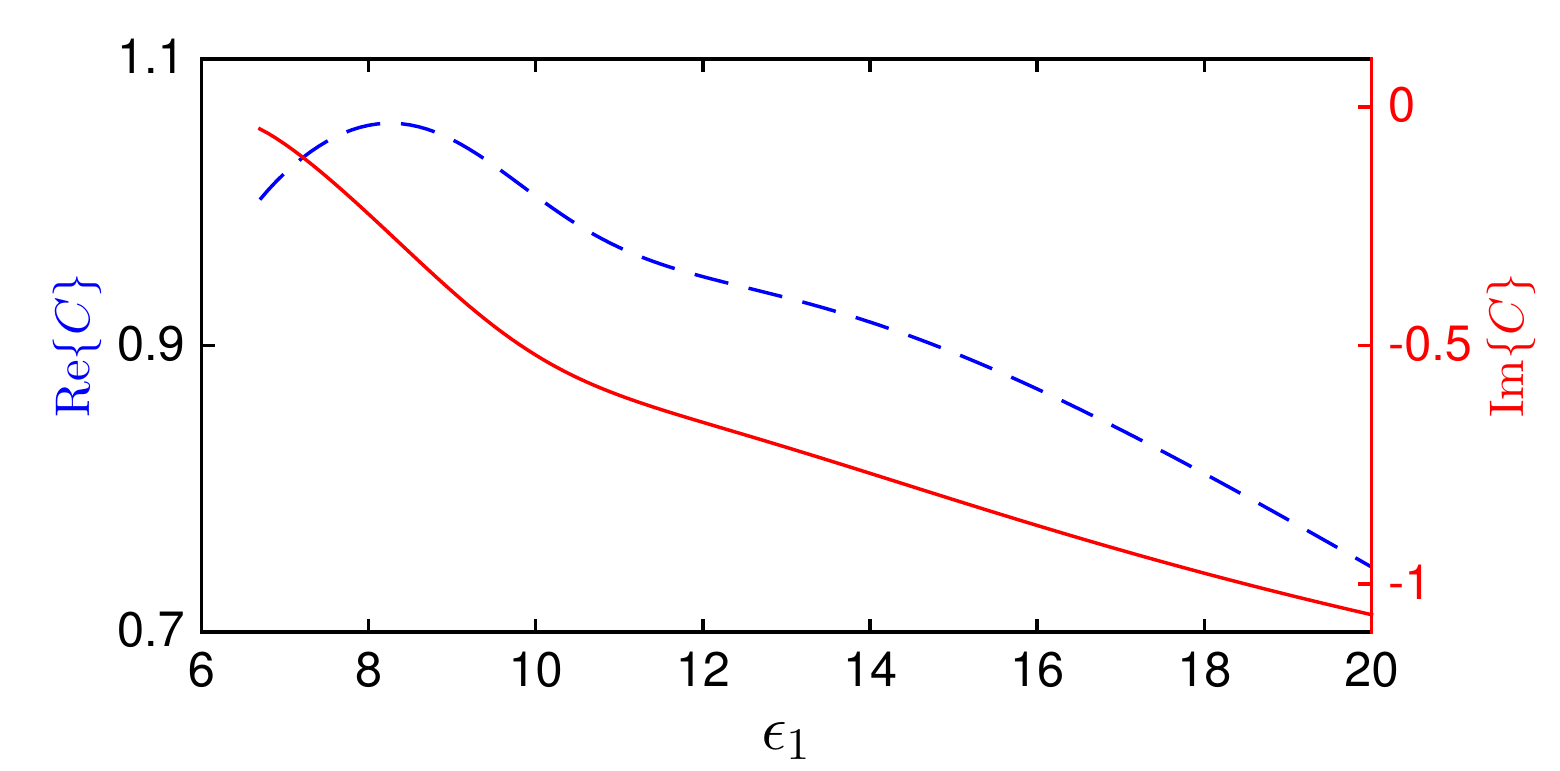}
\caption{The real (dashed blue curve) and imaginary (solid red curve)
  parts of $C_0$ for a propagating BIC as functions of radius $a$
  (left, for fixed $\epsilon_1 = 10$) and dielectric constant
  $\epsilon_1$ (right, for fixed $a = 0.3L$). Real and imaginary parts
  follow the left and right axes.}
\label{fig:Example1_C}
\end{figure}

A periodic array of circular cylinders is a structure with 
reflection symmetries in both $x$ and $y$. Propagating BICs on such 
structures are usually regarded as unprotected by symmetry, but some of these BICs are robust under symmetry-preserving 
perturbations. That is, if the structure is perturbed without 
destroying  the reflection symmetries, the BIC continues its existence 
with  a slightly different frequency and a slightly different Bloch 
wavenumber \cite{yuan17_2}.  For the circular array, the BIC has an 
existence domain in the parameter space of radius $a$ and dielectric 
constant $\epsilon_1$ \cite{yuan17_1}. The perturbation theory of
section~\ref{perturbation} is supposed to be valid for any BIC on 2D
periodic structures. As a simple check, we calculate coefficient $C_0$
varying one of these two parameters. The results are shown in 
Fig.~\ref{fig:Example1_C}. 
The left and right panels show $C_0$ for a fixed $\epsilon_1=10$ and
fixed $a=0.3L$, respectively. The real and imaginary parts of $C_0$ are
shown according to the left and right vertical axes in each panel. 

Since the second harmonic wave is $O(A/\delta)$, a
smaller nonlinear coefficient actually produces a larger amplitude for
the BIC. This may seem to be counter-intuitive, since the BIC is 
created by a nonlinear process, and a smaller nonlinear
coefficient  implies weaker nonlinear effects. However, the amplitude 
of the BIC only reflects the steady state where a number of 
sub-processes are balanced. The terms $u_1^2$ and $u_2^2$ in the
right hand sides of Eqs.~\eqref{eq:FourWavesModel_2} and 
\eqref{eq:FourWavesModel_4}  correspond to the SHG
processes from $\omega$ to $2\omega$ and from $2\omega$ to $4\omega$,
respectively. Power is converted to the second harmonic wave by the
first SHG process, and removed from it by the second SHG
process. A smaller nonlinear coefficient
reduces the ratio between the amplitudes of $u_4$ and $u_2$, but it
does not reduce the ratio between $u_2$ and $u_1$, since 
the second harmonic wave can accumulate in the BIC. As we
mentioned in section~\ref{perturbation}, the term $\bar{u}_2 u_4$ in
the right hand side of Eq.~\eqref{eq:FourWavesModel_2} must have the
same order as $u_1^2$, therefore, a smaller nonlinear coefficient will
produce a stronger second harmonic wave. If we consider a time-dependent
problem by turning on the incident wave at $t=0$, 
a smaller nonlinear coefficient implies that longer time is needed to
reach the steady state. 

The excitation of a BIC of $O(A/\delta)$ by an incident wave of
amplitude $A$ is only possible when $C_0 \ne 0$. It is important to
note that $C_0 = 0$ for anti-symmetric standing waves on
symmetric periodic structures. Here, the reflection symmetry in 
the periodic direction $y$ is concerned. The dielectric function of the
periodic structure  is assumed to be even in $y$, while the BIC mode
profile $u_*$ is odd in $y$, and the Bloch wavenumber $\beta_*=0$. The
anti-symmetric standing waves are well-known symmetry-protected BICs. 
The linear solution $u_{10}$ is an even function of $y$, since
$\beta_*=0$ implies a normal incident wave. The function $F=F(x,y)$ is
also even in $y$, thus $C_0$ given in Eq.~\eqref{eq:C2} is automatically
zero. For such a case, it can be shown that the leading terms for
second, third and fourth harmonic waves are $O(A \delta^3)$, $O(A
\delta^6)$ and $O(A\delta^9)$, respectively, and it is sufficient to
use the two-wave model. 
For $a=0.3L$ and $\epsilon_1=10$, the periodic array of circular
cylinders has an anti-symmetric standing wave with $\omega_* =
0.4415\,(2\pi c/L)$ and 
$\beta_*=0$. We have solved the four-wave model with same incident
wave amplitude $A = 10^6$\,V/m and nonlinear coefficient $\gamma_1 =
10^{-12}$\,m/V, confirmed that the second harmonic wave is weak, and
the higher harmonic waves are extremely weak.


\section{Conclusion}
\label{Conclusion}

The diffraction of a plane wave by a periodic structure  is the subject
of many theoretical, numerical and experimental studies. Nonlinear
diffraction problems give rise to many interesting wave phenomena and have
useful practical applications. For periodic structures with a second
order nonlinearity,  existing studies are concerned with the SHG
process for which it is sufficient to include only the fundamental and
second harmonic waves. In this paper, we also consider 
periodic structures with a second order nonlinearity, 
but assume that the corresponding linear periodic structure has a BIC and the incident 
wave has half the BIC frequency and a compatible wavevector. It turns
out that the nonlinear diffraction problem under these assumptions is
very different from the standard SHG process. The incident wave may
induce a very strong second harmonic wave
dominated by the BIC, and it also induces higher harmonic waves among
which the fourth harmonic wave cannot be ignored. We derive an explicit
formula for the leading coefficient of the BIC using
a perturbation method, and propose that a coupled system of four
Helmholtz equations (the four-wave model) is sufficiently accurate for
analyzing the nonlinear process. 
Numerical solutions based on a four-wave model are presented and used
to validate the perturbation results.  

The BICs on a periodic structure are guided modes above the
lightline that cannot be excited by incident plane waves if the
structure consists of linear materials. Our study indicates that a
propagating BIC with a very large amplitude  can be excited by a plane
incident wave if the structure has a second order nonlinearity. 
Since the fourth harmonic
wave has the largest power conversion efficiency, it may be possible to
realize the nonlinear process studied in this paper as a useful 
fourth harmonic generation technique.

\end{document}